\documentclass[doublecol,letterpaper]{epl2_edit} 
\usepackage{graphicx}
\usepackage{dcolumn}
\usepackage{bm}
\usepackage{dcolumn}
\usepackage{url}
\usepackage{graphicx}
\usepackage{bm}   
\usepackage{bbm}   
\usepackage{verbatim}
\usepackage{stmaryrd}
\usepackage{amsthm}

\newcommand{\Popt} {P_{\rm opt}}
\newcommand{\DKL}[2] { {\cal D\!}\left[ {#1} \| {#2} \right] }

\title{Information theoretic approach to interactive learning}

\author{Susanne Still}
\institute{     
University of Hawaii at Manoa, ICS Department, Honolulu, HI 96822, USA.
\texttt{sstill@hawaii.edu}
}
\notes{Published in EPL}{85 (2009) 28005}

\abstract{
The principles of statistical mechanics and information theory play an important role in learning and have inspired both theory and the design of numerous machine learning algorithms. The new aspect in this paper is a focus on integrating feedback from the learner. A quantitative approach to interactive learning and adaptive behavior is proposed, integrating model- and decision-making into one theoretical framework. This paper follows simple principles by requiring that the observer's world model and action policy should result in maximal predictive power at minimal complexity. Classes of optimal action policies and of optimal models are derived from an objective function that reflects this trade-off between prediction and complexity. The resulting optimal models then summarize, at different levels of abstraction, the process's causal organization in the presence of the learner's actions. A fundamental consequence of the proposed principle is that the learner's optimal action policies balance exploration and control as an {\em emerging} property. Interestingly, the explorative component is present in the absence of policy randomness, i.e. in the optimal {\em deterministic} behavior. This is a direct result of requiring maximal predictive power in the presence of feedback.}

\begin{document}

\maketitle

\section{Introduction}
The problem of learning a model, or model parameters, from observations obtained in experiments, appears throughout physics and the natural sciences as a whole. The statistical mechanics of learning have been discussed in many contexts \cite{WatkinRauBiehl93, MalzahnOpper02}, such as neural networks, support vector machines \cite{SMSVM, OpperUrbanczik01}, and unsupervised learning via compression \cite{Rose90}. The latter, information theoretic approach essentially views learning as lossy compression -- data are summarized with respect to some relevant quantity \cite{IBN}. This can be an average variance \cite{Rose90}, or any other measure of either distortion \cite{Shannon48} or relevance \cite{IBN}. Applied to time series data, one can show \cite{StillCruEl07a} that if prediction is relevant, then representations are found by this approach that constitute unique sufficient statistics \cite{Crut98d} and which can be interpreted as underlying {\em causal states} \cite{Crut88a} of the observed system. 

However, the role of the observer is not always a passive one, as is assumed in the large majority of work on learning theory (see e.g. \cite{SML, Vapnik95}). In many problems ranging from quantum mechanics, to neuroscience, to animal behavior, the interactive coupling between the observer and the system that is being observed is crucial and has to be taken into account.   

In this paper, an {\em information-theoretic approach to integrated model and decision making} is proposed. As a first step towards a general theory of adaptive behavior, let us ask a simple question: If the goal of a learner is to have as much predictive power as possible, then what is the least complex action policy, and what is the least complex world model that achieve this goal? 

The ability to predict improves the performance of a learner across a large variety of specific behaviors, and is hence quite fundamental, increasing the survival chance of an autonomous agent, or an animal, and the success rate on tasks, independent of the specific nature of the task. Furthermore, a good model of the world must generalize well (see, e.g., \cite{Vapnik95})---in other words, the quality of the learner's world model can be judged by how well it predicts as-yet unseen data. For those reasons prediction is in general crucial for any adaptively behaving entity. Therefore, as a first step, we focus on prediction. To model animal behavior, other constraints, such as energy consumption, are clearly also relevant.

The approach taken here is related to, but different from {\em active learning} (e.g., \cite{MacKay92c, SeungOpperSomp92, dasgupta06, AA06}) and { \em optimal experiment design}, which has found countless applications in physics, chemistry, biology and medicine (\cite{Fedorov}; for more recent reviews see, e.g., \cite{atkinson01, boxhunter05}). These approaches do not usually take feedback from the learner into account. Feedback is modeled more explicitly in {\em reinforcement learning} (RL) \cite{SuttonBarto1998}, but this approach is limited to specific inputs, assuming that the learner receives a reward signal. In contrast to RL, we step back and ask about behavior that is optimal with respect to learning about the environment rather than with respect to fulfilling a specific task. Our approach does not require rewards.

Much of the RL literature assumes that the learner's explorative behavior is achieved by some level of randomness of the behavioral policy \cite{kaelbling1996}. Here we show, in contrast, that if learning and optimal model-making are the goal, then explorative behavior emerges as one component of the optimal policy -- even in the absence of stochasticity: any policy which is optimal with respect to learning maximally predictive models must balance exploration with control, including the optimal {\em deterministic} policy (see Eq. (\ref{det.act})). 

Conceptually, our approach could perhaps be thought of as ``rewarding" information gain and, hence, curiosity. In that sense, it is related to {\em curiosity driven RL} \cite{Schmidhuber1991a}, where internal rewards are given that correlate with some measure of prediction error.\footnote{Our approach is fairly general, and to compare one has to adopt the specific RL setting, which we explore in \cite{StillPrecup}.} However, an important difference of the approach discussed here is that the learner's goal is not to predict future rewards, but rather to behave such that the time series it observes as a consequence of its own actions is rich in causal structure. This, in turn, then allows the learner to construct a maximally predictive model of its environment.

\section{Optimally predictive model and decision making}
Let there be a physical system to be learned, and call it the learner's ``world". A learner in parallel (i) builds a model of the world and (ii) engages in an interaction with the world. The learner's inputs are observations, $x(t)$, of (some aspects of) the world. Observations result in actions, $a(t)$, through a decision process. Actions affect the world and so change future observations. 

Let us assume that the learner interacts with the environment between consecutive observations.\footnote{This sequential setup is useful for the sake of simplicity. However, a real agent continuously acts and senses, and an extension to this more involved case would be interesting.} Let one decision epoch consist in mapping the current ``history", $h$ (specified below), available to the learner at time $t$, onto an action (sequence) $a$ that starts at time $t$ and takes the time $\Delta$ to be executed. The next datum is sensed at time $t+\Delta$. (We assume for simplicity that the times it takes to react and to sense are both negligible.)

The decision function, or {\em action policy} \cite{SuttonBarto1998}, is given by the conditional probability distribution $P(a | h)$.\footnote{Short hand notation: the argument $t$ is dropped. Actions $a$, internal states $s$, futures $z$, and histories $h$ are (possibly multi-valued) random variables with values $A \in {\cal A}$, $S \in {\cal S}$, $Z \in {\cal Z}$, and $H \in {\cal H}$, respectively.} Let the model summarize historical information, using internal states $s$, via the probabilistic map $P(s | h)$.
The model and the policy depend upon each other, but histories are mapped {\em independently} onto (i) internal states (using the model $P(s|h)$), and (ii) action sequences (using the policy $P(a|h)$). Hence, actions and internal states are conditionally independent, if the history $h$ is given: 
\begin{equation}
P(s, a | h) = P(s|h)P(a|h).  
\label{cond.ind}
\end{equation}
The ``internal state" does not change the statistics of the environment, but rather serves as an internal observer. The feedback due to the actions, however, changes the statistics of the environment. The action policy contains a model in the sense that if a large group of histories share the same optimal action, then the action can be viewed as a compressed representation of this ``history-cluster". 

The learner uses the current state, $s(t)$, together with knowledge of the action, $a(t)$, to make probabilistic predictions of future observations, $z(t)$, of length $\tau_f$:\footnote{Future observations, $z(t)$, are given by the signal $x(t')$ on the interval \mbox{$t' \in [t +\Delta , t+\Delta+\tau_f]$}, where $\Delta$ is the duration of the intervention given by the action, or the sequence of actions, initiated at time t, $a(t)$. The learner is interested in understanding how one intervention changes the future. The action choice does depend on past actions, if they are included in the learner's history, $h(t)$. However, planning of consecutive future actions is not discussed here, but an extension would be desirable. The notation $\langle \cdot \rangle_{P}$ denotes the average taken over $P$.} 
\begin{equation} 
P(z|s,a) = {1\over P(s,a)} \left\langle P(z|h,a) P(a|h) P(s|h)\right\rangle_{P(h)}.
\label{pzgsa}
\end{equation} 
$P(z|h,a)$ and $P(h)$ are (for the moment) assumed to be known. A history always includes the current observation, $x(t)$. Beyond this, it may include a record of prior observations reaching some length $\tau_p$ into the past, and also previous internal state and action(s). Lengths of the internal records of past observations and past actions are assumed given by the learner's storage capacity.

The problem of interactive learning then is to choose a model and an action policy, which are optimal in that they maximize the learner's ability to predict the world, while being minimally complex. 

We measure the learner's predictive ability by the mutual information \cite{Shannon48} that the internal state, {\em in the presence of the action}, contains about the future: 
\begin{equation} 
I[\{s,a\};z] = \left\langle \log{\left[ {P(z|s,a) \over P(z)} \right]}\right\rangle_{P(z,s,a)}.
\label{pred.inf}
\end{equation} 
The quantity $I[\{s,a\};z] = H[z] - H[z|s,a]$ measures the reduction in the uncertainty about the future (entropy $H$), when state and action are known. It is zero if the future is independent of $s$ and $a$. It is maximal if the knowledge of $s$ and $a$ eliminates all uncertainty about the future ($H[z|s,a] = 0$).

Simple models and simple action policies come at a lower coding cost, quantified by the coding rates $I[s;h]$ and $I[a;h]$, respectively. The notion that the simplest possible model is preferable is deeply rooted in our culture. William of Ockham is frequently cited on this matter, which is known as ``Ockham's razor". In the same vein, out of two action policies which yield the same value of the objective, Eq. (\ref{pred.inf}), one would choose the simpler policy, as there is no reason to implement a more complex policy which takes more memory.

The interactive learning problem is solved by maximizing $I[\{s,a\};z]$ over $P(s|h)$ and $P(a|h)$, under constraints that select for the simplest possible model and the most efficient policy, respectively, in terms of smallest complexity measured by the coding rate. Less complex models and policies result in less predictive power. This trade-off can be implemented using Lagrange multipliers, $\lambda$ and $\mu$. Following the spirit of \emph{rate distortion theory} \cite{Shannon48}, and, more closely related, the \emph{information bottleneck method} (IB) \cite{IBN}, one can then calculate the best possible solution at each value of the Lagrange multipliers. The optimization problem for interactive learning is given by:
\begin{equation}
\begin{array}{cl}
\max & \bigl( I[\{s, a\}; z] - \lambda I[s; h] - \mu I[a; h] \bigr) \\
P(s | h) & \\
P(a | h) &
\end{array}
\label{obj.fkt.2}
\end{equation} 
The two constraints are taken into account individually, rather than as a sum,\footnote{\mbox{$I[\{s,a\};h] + I[s;a] = I[a;h] + I[s;h]$}, because of Eq. \ref{cond.ind}. $I[\{s,a\};h]$ is the coding rate of the learner's full behavior -- consisting of both the internal state, $s$, and the action sequence, $a$. $I[s;a]$ measures the redundancy, which should be minimized together with the coding rate.}
so that their relative importance can be adjusted.
Think, for example, about a robotic multi-agent system in which robots communicate their internal states to each other. Limited communication channel capacity may force them to produce compact internal representations, but the complexity of the action policy that each individual can implement does not have to be equally constrained. 

The trade-off parameters $\lambda$ and $\mu$ parameterize families of optimal models and policies, respectively, constituting those models and policies that have maximal predictive power at fixed complexity.  An analogy to statistical mechanics is useful to guide intuition \cite{Rose90}, and relates $\lambda$ and $\mu$ to temperature -- they control the "fuzziness" of the maps that assign histories to states and actions, respectively. This approach also relates the distortion function to the energy function of a corresponding physical system and the normalization constant to the partition function. 

\subsection{Optimal action policies} 
The action policies that solve optimization problem, Eq. (\ref{obj.fkt.2}), are given by
\begin{equation}
\Popt (a | h) = {P(a) \over Z_{\rm A}(h,\mu)}  e^{ - {1 \over \mu} E_{\rm A} (a,h)}
\label{optimalpolicy} 
\end{equation}
with the energy function
\begin{eqnarray}
E_{\rm A} (a,h) &=&\left\langle \DKL{P(z|h,a)}{P(z|s,a)} \right\rangle_{P(s|h)} \nonumber \\
&&- \DKL{P(z|h,a)}{P(z)}, 
\label{energy.action}
\end{eqnarray}
and the partition function
\begin{equation}
Z_{\rm A}(h,\mu) = \bigl\langle e^{ - {1 \over \mu} E_{\rm A} (a,h)} \bigr\rangle_{P(a)}.
\label{Z.action}
\end{equation}
$\DKL{p}{q} = \langle \log [p/q] \rangle_{p}$ denotes the {\em relative entropy}, or {\em Kullback--Leibler divergence} between distributions $p$ and $q$. 
Equations (\ref{optimalpolicy})-(\ref{Z.action}) must be solved self-consistently, together with Eq. (\ref{pzgsa}) and
\begin{eqnarray}
P(a) &=& \left\langle P(a|h) \right\rangle_{P(h)}, \\
\label{pa}
P(z) &=& \left\langle \left\langle P(z|h,a) \right\rangle_{P(a|h)} \right\rangle_{P(h)}.
\label{pz}
\end{eqnarray} 
To derive this result (Eqs. (\ref{optimalpolicy})-(\ref{Z.action})), one calculates $I[\{s,a\};z]$, using Eq. (\ref{pzgsa}), and the functional derivative of Eq. (\ref{obj.fkt.2}) w.r.t. $P(a|h)$. Individual nonzero contributions are given by:\footnote{Terms constant in $a$ are omitted, because in the solution they are absorbed into $Z_{\rm A}$.}
\begin{eqnarray}
\frac{\delta I[\{s,a\};z]}{\delta P(a|h)} &=& P(h) \left\langle \left\langle \log{\left[ \frac{P(z|s,a)}{P(z)} \right]} \right\rangle_{P(z|h,a)} \right\rangle_{P(s|h)} \nonumber \\
&=& P(h) \DKL{P(z|h,a)}{P(z)} \\
&& - P(h) \left\langle \DKL{P(z|h,a)}{P(z|s,a)} \right\rangle_{P(s|h)} \nonumber \\
\frac{\delta I[a;h]}{\delta P(a|h)} &=& P(h) \log{ \left[ \frac{P(a|h)}{P(a)} \right]} ~.
\end{eqnarray}

Observe that the most likely action is that of minimum energy (see Eq. (\ref{optimalpolicy})). The first term in the energy function, Eq. (\ref{energy.action}),
\begin{equation}
\left\langle \DKL{P(z|h,a)}{P(z|s,a)} \right\rangle_{P(s|h)}
\end{equation}
is smaller for actions that will, on average, make the conditional future distribution $P(z|h,a)$ as {\em similar} as possible to the distribution that is {\em predicted} by the learner's internal state, $P(z|s,a)$. The average is taken over the model $P(s|h)$. This term selects for actions that bias the future towards what the learner predicts -- it is therefore related to the {\em control} that the learner can exert on the world.

The second (negative) term in Eq. (\ref{energy.action})
\begin{equation}
- \DKL{P(z|h,a)}{P(z)}
\end{equation}
selects for actions that will make the conditional future distribution $P(z|h,a)$ as {\em different} as possible from the average $P(z)$. The term embodies a preference for actions that bias towards an uncommon future distribution -- it is related to {\em exploration} and causes the learner to perturb the world away from the average.

This shows that at the root of interactive learning there is a competition between exploration and control, which arises as a fundamental consequence of the proposed optimization principle: Exploration and control have to be {\em balanced} in the optimal action policy to result in maximal predictive power. 

\subsection{Optimally predictive models}
The family of models that solve optimization problem Eq. (\ref{obj.fkt.2}), is given by \footnote{The derivation is similar to that for Eq. (\ref{optimalpolicy}) and follows \cite{IBN}. Individual contributions to the functional derivative w.r.t. $P(s|h)$ are (ignoring constant terms):\\ $
\frac{\delta I[\{s,a\};z]}{\delta P(s|h)} = - P(h) \left\langle \DKL{P(z|h,a)}{P(z|s,a)} \right\rangle_{P(a|h)}$ and\\
$\frac{\delta I[s;h]}{\delta P(s|h)} = P(h) \log{ \left[ \frac{P(s|h)}{P(s)} \right]}$.}
\begin{equation}
\Popt (s | h) = {P(s) \over Z_{\rm S}(h,\lambda)}  e^{ - {1 \over \lambda} E_{\rm S} (s,h)}
\label{optimalmodel} 
\end{equation}
with 
\begin{equation}
E_{\rm S} (s,h) = \left\langle \DKL{P(z|h,a)}{P(z|s,a)} \right\rangle_{P(a|h)}  
\label{energy.model}
\end{equation}
and
\begin{equation}
Z_{\rm S}(h,\lambda) = \bigl\langle e^{ - {1 \over \lambda} E_{\rm S} (s,h)} \bigr\rangle_{P(s)} ~.
\end{equation}
These equations must be solved self-consistently, together with Eq. (\ref{pzgsa}) and
\begin{equation}
P(s) = \left\langle P(s|h) \right\rangle_{P(h)}.
\label{ps}
\end{equation}

The most likely state minimizes the relative entropy between the actual, $P(z|h,a)$, and the predicted, $P(z|s,a)$, conditional future distribution (see Eqs. (\ref{optimalmodel}) and (\ref{energy.model})), averaged over the action policy $P(a|h)$. The internal states thus capture the effect that the history has on the probability distribution over futures, under a given action policy. In that sense, the optimal model reflects the causal structure of the underlying process. 

Altogether, Eqs. (\ref{optimalpolicy}) and (\ref{optimalmodel}), must be solved self consistently (together with Eqs. (\ref{pzgsa}), (\ref{energy.action})-(\ref{pa}), and (\ref{energy.model})-(\ref{ps})) to yield the model that is optimally predictive under the optimal policy (and vice versa). This can be done iteratively, resulting in an algorithm that is similar to the IB algorithm \cite{IBN}. This new algorithm, however, includes a feedback loop, due to actions.\footnote{Details about the algorithm are given in \cite{StillBialek06}, where examples are also discussed. An extension will be published elsewhere.}

With increasing $\lambda$, the level of abstraction of the model increases, as less detail is kept. In the high temperature limit, $\lambda \rightarrow \infty$, all possible histories are effectively represented by the same internal state.\footnote{As $\lambda \rightarrow \infty$, $P(z|s,a)$ is the same for all states $s$: As $\lambda \rightarrow \infty$, $\Popt (s | h) \rightarrow P(s)$, see Eq. (\ref{optimalmodel}), and with that \mbox{$P(s,a) = \langle P(s,a|h) \rangle_{P(h)} = \langle P(s|h) P(a|h) \rangle_{P(h)} \rightarrow P(s) P(a)$}, and \mbox{$P(z|s,a) \rightarrow {1\over P(s) P(a)} \left\langle P(z|h,a) P(a|h) P(s)\right\rangle_{P(h)} = P(z|a)$, $\forall s$}; see Eqs. (\ref{cond.ind}) and (\ref{pzgsa}).}

\subsection{Deterministic models and decisions} 
In the low temperature limit ($T \rightarrow 0$; $T \in \{ \lambda, \mu \}$), the distributions in Eqs. (\ref{optimalpolicy}) and (\ref{optimalmodel}) become deterministic mappings. 
To see this, let us use the discrete random variable $y \in \{a,s\}$, and let $E(y,h)$ denote the value of the energy function $E_{\rm A}$, if $y=a$, and $E_{\rm S}$, if $y=s$. Furthermore, define the functions
$y^*(h) := {\rm arg}\min_y E(y,h)$ 
and ${\cal E}(y,h) := E(y,h) - E(y^*(h),h) \geq 0$. Now, we can write the conditional distribution for the most likely value $y^*(h)$ as \\
\begin{eqnarray}
P(y = y^*(h)|h) &=& {P(y = y^*(h)) \over Z(h, T)} e^{- {1 \over T} E(y^*(h),h)} \nonumber\\
&=&\left( 1 + \sum_{y \neq y^*(h)} {P(y) \over P(y^*(h))} e^{-{1 \over T}{\cal E}(y,h)} \right)^{-1} ~.
\end{eqnarray}
Since ${\cal E}(y,h)$ is positive, the sum goes to zero as $T \rightarrow 0$ (assuming that $P(y^*(h)) > 0$).
As a consequence, we have \mbox{$P(y = y^*(h)|h) = 1$} and, due to normalization, the optimal mapping becomes deterministic: 
\mbox{$P_{T \rightarrow 0}(y|h) = \delta_{y y^*(h)}$}, where $\delta$ denotes the Kronecker-Delta.

For a {\em deterministic model}, specified by \mbox{$P_{\lambda \rightarrow 0}(s|h) = \delta_{s s^*(h)}$}, this means that a history $h$ is assigned with probability one to the state $s = s^*(h)$ which minimizes the energy function $E_{\rm S}(s,h)$, Eq. (\ref{energy.model}):
\begin{equation}
s^*(h) = {\rm arg}\min_s \left\langle \DKL{P(z|h,a)}{P(z|s,a)} \right\rangle_{P(a|h)}.
\label{s_star}
\end{equation}
Note that without constraints on the cardinality of the state space, one can always ensure that this minimum is zero: $E_{\rm S}(s^*(h),h) = 0$. This fact then implies that the predicted information, $I[\{s, a\}; z]$, reaches its maximum at the optimal deterministic model, $I[\{s^*, a\}; z]$. 

The maximum is given by the predictive information of the time series, in the presence of the learner's actions: $I[\{s^*, a\}; z] = I[\{h, a\}; z]$.\footnote{$I[\{s, a\}; z] = I[\{h, a\}; z] - \langle E_{\rm S}(s,h) \rangle_{p(s,h)}$. The second term vanishes for the optimal deterministic model. It becomes $\langle E_{\rm S}(s^*(h),h) \rangle_{p(h)} = 0$.} The optimal policy now maximizes this quantity, at fixed $I[a,h]$. This illustrates that the optimal policy makes as much information as possible available to be summarized by the model, at fixed policy complexity. 

Action policies become increasingly random with increasing $\mu$ -- the learner's reactions become less specific responses to the history. In the other limit, as the complexity constraint is relaxed by letting the parameter $\mu$ approach zero, one finds the optimal {\em deterministic} policy $a^*(h)$\footnote{$a^*(h)$ is given by Eq. \ref{det.act}.} which maximizes the predictive information of the time series, in the presence of the actions. 

The special case is of particular interest in which the learner produces {\em deterministic} maps $s^*(h)$ and $a^*(h)$, which maximize the predictive power, Eq \ref{pred.inf}. The optimal deterministic model maps a history $h$ to the internal state 
\begin{equation}
s^*(h) := {\rm arg}\min_s \DKL{P(z | h, a^*(h))}{P(z | s, a^*(h))}. 
\label{opt.s.det}
\end{equation}
Assuming that there are no constraints on the cardinality of the state space, this map partitions the space of histories in a way that is similar to the {\em causal state partition} of \cite{Crut88a}. One can show \cite{StillCruEl07a} that if actions are not considered (passive time series modeling), then the passive equivalent of Eq. (\ref{opt.s.det}) exactly recovers the causal state partition of \cite{Crut88a}. Causal states are unique and minimal sufficient statistics -- constituting a meaningful representation of the underlying process \cite{Crut98d}. 

The partition specified by Eq. (\ref{opt.s.det}) allows for an extension of the causal state concept to {\em interactive} time series modeling: here the space of histories is partitioned such that all histories, $h \in {\cal H}_{s} \subset {\cal H}$, that are mapped to the same causal state, $s$, are {\em causally equivalent under the optimal action policy}, $a^*(h)$; meaning that their conditional future distributions $P(z |h, a^*(h))$ are the same. 

This grouping of histories results in an equivalence class that is controlled by the action policy: under any action policy, $A(h)$ (where the map $A: h \mapsto a$ is a deterministic policy), two histories $h$ and $h'$ are equivalent with respect to their effect on the future, $z$, if $P(z |h, A(h)) = P(z |h', A(h'))$. The resulting partition, ${\cal S}_A$, of the history space into causal states depends on the action policy, $A$. The choice of the policy determines the nature of the time series which is produced by the system {\em coupled} to the observer through the actions. Note that there could be different action policies $A' \neq A$, which result in coupled systems with the same underlying causal state partition ${\cal S}_A = {\cal S}_{A'}$. The policy $A = a^*$ is the deterministic policy that creates the coupled world-observer system that can be predicted most effectively by a causal model. 

Optimal deterministic decisions for actions are made according to the rule
\begin{eqnarray}
a^*(h) := {\rm arg}\min_a \bigl[&&\left\langle \DKL{P(z|h,a)}{P(z|s,a)} \right\rangle_{P(s|h)}  \nonumber \\ 
&& - \DKL{P(z|h,a)}{P(z)} \bigr].
\label{det.act}
\end{eqnarray}
It is important to note that the term related to exploration (second term) persists in the optimal deterministic action policy, Eq. (\ref{det.act}). This is in direct contrast to ``Boltzmann exploration", commonly used in RL \cite{SuttonBarto1998}. There, exploration is implemented as policy randomization by softening of the optimal, deterministic policy (optimal in an RL sense by maximizing expected future reward). We have shown here, however, that to create data which allows for optimally predictive modeling, an exploratory component must be present even in the optimal {\em deterministic} policy. In our framework, exploration is hence an emerging behavior, and it is {\em not} the same as policy randomization.

\subsection{Probability estimates and finite sampling errors}
So far, we have assumed $P(z | h, a)$ and $P(h)$ to be known. However, in practice, they may have to be estimated from the observed time series. Hence there could be a bias towards overestimating $I[\{s,a\};z]$ due to finite sampling errors in the probability estimates. This may result in over-fitting. The accuracy of the estimates depends on the data set size, $N$. One can counteract finite sampling errors, using an approximate error correction method, such as discussed in \cite{StillBialek2004}. This method has already been applied successfully to predictive inference in the absence of actions \cite{StillCruEl07a} and it can also be applied in the presence of actions. 

\subsection{Time dependent on-line learning procedure}
In \cite{StillBialek2004}, we calculated bounds on the smallest temperature, $T^*(N)$, allowable before over-fitting occurs. This value depends on the data set size $N$. In the interactive learning setup, the data set size grows linearly with time. One can implement an algorithmic annealing procedure, similar to the one in \cite{Rose90}, but different in that the temperature is kept fixed at each time step and then changes over time with growing data set size. This captures the intuition that a learner may allow itself to model an increasing amount of detail the longer it has observed the world. The temperatures in each time step are set to (upper bounds on) the values $\lambda^*(t)$ and $\mu^*(t)$, below which over-fitting would occur. Since these can be calculated, an annealing {\em rate}, as used in deterministic annealing \cite{Rose90}, is not necessary. The work in \cite{StillBialek2004} directly provides a bound on $\lambda^*(t)$ and could be extended to calculate a bound on $\mu^*(t)$. Tighter bounds or an exact calculation of $T^*$ would also be desirable.
\vglue -2cm
\subsection{Possible extension to multi-agent systems}
When multiple agents observe and interact with an environment, they often exhibit emerging co-operative behavior. Understanding the emergence of such co-operative strategies is an active field of research. In order to utilize our approach for the study of this phenomenon, we have to distinguish (i) the agents' available sensory input and (ii) whether there is communication between agents. In the simplest case, each of the agents has access only to data from the environment. Then each agent can be modeled exactly as we have outlined here, and all coupling happens implicitly, through the environment. Communication of internal states and/or the observation (or communication) of each others actions, however, means that the {\em other agents'} internal states and/or actions, respectively, must be included in each agent's input (history $h$). Furthermore, if agents try to learn about each others behavior, then we need to include the other agents' future actions into the data which ought to be predicted (future $z$). A detailed exploration of multi-agent learning has to be left for future research.
\vglue -2cm
\section{Summary} This paper has proposed an information-theoretic approach to a quantitative understanding of interactive learning and adaptive behavior by means of optimal predictive modeling and decision making. A simple optimization principle was stated: use the least complex model and action policy which together provide the learner with the largest predictive ability. A fundamental consequence of this principle is that the optimal action policy finds a balance between exploration and control. This is a direct consequence of optimal prediction in the presence of feedback due to the learner's actions.
The theory developed here is general in that it makes no assumptions about the detailed structure of the underlying process that generates the data, and thus is not restricted to specific model classes. 
\vglue -2cm
\section{Acknowledgments} I am deeply grateful to W. Bialek who contributed significantly to the ideas expressed in this paper. I thank J. P. Crutchfield and D. Precup for many helpful discussions and comments on the manuscript, and M. Berciu, L. Bottou, I. Nemenman, B. A. Pearlmutter and C. Watkins for very useful discussions. 

\bibliography{interactive_learning_references}

\end{document}